\documentclass[prc,preprint]{revtex4}

\usepackage{graphicx}

\begin{document}
\title{Influence of spin polarizability on liquid gas phase transition
in the nuclear matter}

\author{Z. Rezaei$^{1}$, M. Bigdeli$^{2}$ and G. H. Bordbar$^{1,3}$}

\address{$^{1}$Department of Physics,
Shiraz University, Shiraz 71454, Iran\\
$^{2}$Department of Physics, University of Zanjan,
Zanjan 45371-38791, Iran\\
 $^{3}$Center for Excellence in Astronomy and Astrophysics (CEAA-RIAAM)-Maragha,
P.O. Box 55134-441, Maragha 55177-36698, Iran}

\begin{abstract}
In this paper, we investigate the liquid gas phase transition for the spin polarized nuclear matter.
Applying the lowest order constrained variational (LOCV) method, and using two microscopic potentials, $AV_{18}$ and $UV_{14}$+TNI, we calculate the free energy, equation of state, order parameter, entropy, heat capacity and compressibility
to derive the critical properties of spin polarized nuclear matter. Our results indicate that for
the spin polarized nuclear matter, the second order phase transition takes place at lower temperatures
with respect to the unpolarized one. It is also shown that the critical temperature of our spin polarized nuclear matter with a specific value of spin polarization parameter is in good agreement with the experimental result.
\end{abstract}
\maketitle
\noindent Keywords: {polarized nuclear matter; liquid gas phase transition.
%%%%%%%%%%%%%%%%%%%%%%%%%%%%%%%%%%%%%%%%%%%%%%%%%%%%%%%%%%%%%%%%%%%%%%%%%%%%%%%%%%%%%%%%%%%%%%%%%%%%%%%%%%%%%%%%%%%%%%%%%%%%%%%%%%%%%%%%%%%%%

\section{Introduction}

Through the formation of protoneutron stars and in the relativistic heavy ion collisions,
the symmetric nuclear matter can reach the high temperatures about $20-50 \ MeV$ \cite{Chapline,Camenzind}.
Therefore, liquid gas phase transition in nuclear matter may occur.
To investigate the nature of liquid gas phase transition and
the properties of critical point, many experimental \cite{Finn,Curtin,Pochodzalla,Natowitz} and theoretical \cite{Friedman,Huber,Malheiro,Baldo,Zuo,Rios,Rios2010,Zhang}
studies have been done.
The extracted value of critical temperature from the heavy ion collisions
is $T_c\simeq16.6\ MeV$ \cite{Natowitz}. However, the critical temperature driven by theoretical
methods depends on the applied models and interactions.
Using the variational method based on
the  $UV_{14}$ two-body nucleon-nucleon (NN) interaction including a phenomenological three-body force (TNI),
the critical temperature has been gotten about $17.5\ MeV$ \cite{Friedman}.
In the frame of relativistic
Brueckner-Hartree-Fock (BHF) theory and applying
the potentials given by Brockmann and Machleidt, it has been shown that
the critical temperatures are smaller than those of the nonrelativistic
investigations \cite{Huber}.
Applying different versions of scalar derivate coupling suggested by Zimanyi and
Moszkowski, the critical temperature has been obtained in the range of $13.6-18.3\ MeV$ \cite{Malheiro}.
Using the Bloch-De Dominicis diagrammatic expansion
and employing the $AV_{14}$ as the bare NN interaction,
without and with a phenomenological three-body force,
the gotten critical temperatures are $T_c\simeq21\ MeV$ and $T_c\simeq20\ MeV$, respectively
\cite{Baldo}.
In the framework of finite temperature BHF approach extended
to include the contribution
of a microscopic three-body force, it has been shown that the three-body force reduces
the critical temperature  from  $T_c\simeq16\ MeV$ to $13\ MeV$ \cite{Zuo}.
It has been seen that the critical temperature for the CDBONN potential in the BHF approximation ($T_c\simeq 23.3\ MeV$) is larger than the critical temperature for the same potential but in the Self-Consistent Greens Functions approach (SCGF) ($T_c\simeq18.5\ MeV$) \cite{Rios}.
In this reference, it has also been shown that for $AV_{18}$ potential, the critical temperature is lower than for the
CDBONN potential \cite{Rios}.
Employing the self-consistent Hartree-Fock approach using different mean-field interactions of the Skyrme and the
Gogny types, it has been indicated that the  critical temperatures span
a wide range of values, from $T_c\simeq14$ to $23\ MeV$ showing
the effective interaction dependence of the critical properties \cite{Rios2010}.
In addition, the predicted value of
critical temperature for the density-dependent relativistic mean-field models is about
$15.7\ MeV$ \cite{Zhang}.

Previously, we
have investigated the liquid gas phase transition of spin unpolarized asymmetrical nuclear matter, using the lowest order constrained variational (LOCV) method \cite{Bordbar01}. The dependence of the critical quantities on the isospin
polarization of nuclear matter has been discussed.
In the present work, we are interested to consider the effect of spin polarizability of symmetric nuclear matter on the liquid gas phase transition applying the LOCV calculation
using the microscopic potentials.

%%%%%%%%%%%%%%%%%%%%%%%%%%%%%%%%%%%%%%%%%%%%%%%%%%%%%%%%%%%%%%%%%%%%%%%%%%%%%%%%%%%%%%%%%%%%%%%%%%%%%%%%%%%%%%%%%%%%%%%%%%%%%%%%%%%%%%%%%%%%%

\section{Finite Temperature Calculations for Spin Polarized Nuclear Matter with the LOCV Method}

We consider a system of $A$ interacting nucleons. This system is
composed of spin-up and spin-down nucleons with densities $\rho_{n}^{(+)}$, $\rho_{p}^{(+)}$,  $\rho_{n}^{(-)}$ and
$\rho_{p}^{(-)}$. Labels ($+$) and ($-$) are used for the spin-up and spin-down neutrons ($n$) and protons ($p$),
respectively. The total densities for neutrons
($\rho_n$), protons ($\rho_{p}$) are given by,
\begin{eqnarray}
     \rho_{n}&=&\rho_{n}^{(+)}+\rho_{n}^{(-)},\nonumber\\
     \rho_{p}&=&\rho_{p}^{(+)}+\rho_{p}^{(-)}.
\end{eqnarray}
The total density of system is
\begin{eqnarray}
     \rho&=&\rho_{n}+\rho_{p}.
 \end{eqnarray}
In the case of symmetric nuclear matter, we have $\rho_n=\rho_p$.
The spin polarization parameters which describe the spin
asymmetry of the system are defined as,
\begin{eqnarray}
      \delta_{p}=\frac{\rho_{p}^{(+)}-\rho_{p}^{(-)}}{\rho_{p}} , \ \ \
      \delta_{n}=\frac{\rho_{n}^{(+)}-\rho_{n}^{(-)}}{\rho_{n}}.
 \end{eqnarray}

To obtain the macroscopic properties of this system, we should
calculate the total free energy per nucleon, $F$,
\begin{eqnarray}\label{free}
       F=E-{\cal T}(S_n^{(+)} +S_p^{(+)}+S_n^{(-)}+S_p^{(-)}),
 \end{eqnarray}
where $E$ is the total energy per nucleon and $S_j^{(i)}$ is the entropy per
nucleon corresponding to the isospin and spin projection $j,i$, respectively,
 \begin{eqnarray}
 S_j^{(i)}(\rho,T)&=&-\frac{1}{A}\sum _{k}
 \left\{\left[1-n_j^{(i)}(k,{\cal T},\rho_j^{(i)})\right]\ln \left[1-n_j^{(i)}(k,{\cal
 T},\rho_j^{(i)})\right]\right.\nonumber \\&& \left.
 +n_j^{(i)}(k,{\cal T},\rho_j^{(i)}) \ln  n_j^{(i)}(k,{\cal T},\rho_j^{(i)})\right\}.
\end{eqnarray}
In above equation,  $n_j^{(i)}(k,{\cal T},\rho_j^{(i)})$ is the Fermi-Dirac
distribution function,
\begin{eqnarray}
n_j^{(i)}(k,{\cal T},\rho_j^{(i)})=\frac{1}{\exp\left({\left[\epsilon_j^{(i)}
(k,{\cal T},\rho_j^{(i)})-\mu_j^{(i)}({\cal T},\rho_j^{(i)})\right]/k_B{\cal T}}\right)+1
},
\end{eqnarray}
where $\epsilon_j^{(i)}$ is the single particle energy of a nucleon, and $\mu_j^{(i)}$ is
the chemical potential which is determined at any adopted value of the temperature $\cal T$ and number density $\rho_j^{(i)}$
 by applying the following constraint,
 \begin{eqnarray}\label{chpt}
 \sum _{k}
 n_j^{(i)}(k,{\cal T},\rho_j^{(i)})=N_j^{(i)}.
 \end{eqnarray}
%
%and $\epsilon^{(i)}$ is the single particle energy of a nucleon.
In our formalism, the single particle energy of nucleons with
momentum $k$
% and spin projection $i$ is
approximately is written in
terms of the effective mass as follows \cite{apv,dapv}
 \begin{eqnarray}
 \epsilon_j^{(i)}(k,{\cal T},\rho_j^{(i)})=
 \frac{\hbar^{2}{k^2}}{2{m_j^{*}}^{(i)}(\rho,{\cal T})}+U_j^{(i)}({\cal
 T},\rho_j^{(i)}),
\end{eqnarray}
where $U_j^{(i)}({\cal
 T},\rho^{(i)})$ is the momentum independent single particle potential.
In fact, we use a quadratic approximation for the single particle
potential incorporated in the single particle energy as a momentum
independent effective mass.  We introduce the
effective masses, $m_j^{{*}{(i)}}$, as variational parameters
\cite{Friedman,Bordbar78}. We minimize the free energy with respect to
the variations in the effective masses and then we obtain the
chemical potentials and the effective masses of the spin-up and
spin-down nucleons at the minimum point of the free energy. This
minimization is done numerically.

For calculating the total energy of polarized symmetric nuclear matter, we
use the LOCV method.
We choose a trial many-body wave function of the form
\begin{eqnarray}
     \psi=\cal{F}\phi,
 \end{eqnarray}
where $\phi$ is the uncorrelated ground state wave function
(the Slater determinant of plane waves) of $A$ independent
nucleons and ${\cal F}={\cal F}(1\cdots A)$ is an appropriate
A-body correlation operator which can be replaced by a Jastrow
form i.e.,
\begin{eqnarray}
    {\cal F}={\cal S}\prod _{i>j}f(ij),
 \end{eqnarray}
where ${\cal S}$ is a symmetrizing operator.
Now, we consider the cluster expansion of energy functional up
to the two-body term \cite{clark},
 \begin{eqnarray}\label{tener}
           E([f])=\frac{1}{A}\frac{\langle\psi|H\psi\rangle}
           {\langle\psi|\psi\rangle}=E _{1}+E _{2}\cdot
 \end{eqnarray}
For the hot nuclear matter, the one-body term $E _{1}$ is
 \begin{eqnarray}
               E_1=\sum_{j=n,p}\sum_{i=+,-}\sum _{k}
               \frac{\hbar^{2}{k^2}}{2m_j}n_j^{(i)}(k,{\cal
               T},\rho_j^{(i)}).
                \end{eqnarray}
The two-body energy $E_{2}$ is
\begin{eqnarray}
    E_{2}&=&\frac{1}{2A}\sum_{ij} \langle ij\left| \nu(12)\right|
    ij-ji\rangle,
\end{eqnarray}
 where
\begin{eqnarray}
 \nu(12)=-\frac{\hbar^{2}}{2m}[f(12),[\nabla
_{12}^{2},f(12)]]+f(12)V(12)f(12).
\end{eqnarray}
In above equation, $f(12)$ and $V(12)$ are the two-body
correlation function and potential.
In our calculations, we use the microscopic potentials ($AV_{18}$ \cite{Wiringa} and $UV_{14}$+TNI \cite{pand}).
The two-body correlation function, $f(12)$, which is induced by the strong force is given by $f(12)=\sum^3_{k=1}f^{(k)}(r_{12})P^{(k)}_{12},$ where $P^{(k)}_{12}$ has been given in
Ref.~\cite{Bordbar80}.
Using this two-body correlation function and
the microscopic potentials, after doing some algebra, we get an equation
for the two-body energy.
In the next step, we minimize the two-body energy with respect to the
variations in the functions $f^{(i)}$ subject to the
normalization constraint, $\frac{1}{A}\sum_{ij}\langle ij| h_{S_{z}}^{2}
-f^{2}(12)| ij\rangle _{a}=0$, \cite{Bordbar80}.
In the case of polarized symmetric nuclear matter, the Pauli
function $h_{S_{z}}(r)$ is as follows \cite{Bordbar80}
\begin{eqnarray}
h_{S_{z}}(r)=
\left\{%
\begin{array}{ll}
\left[ 1-\frac{1}{2}\left( \frac{\gamma^{(i)}(r)
       }{\rho}\right) ^{2}\right] ^{-1/2} & ;\ \hbox{$S_{z}=\pm 1$} \\
    1 & ;\ \hbox{$S_{z}= 0$}
\end{array}%
\right.
\end{eqnarray}
 where
\begin{eqnarray}
\gamma^{(i)}(r)=\frac{1}{\pi^{2}}\int n^{(i)}(k,{\cal
T},\rho^{(i)})J_{0}(kr)k^2dk .
 \end{eqnarray}
From the minimization of the two-body cluster energy, we get a set of
coupled and uncoupled differential equations \cite{Bordbar80}. By
solving these differential equations, we can obtain correlation
functions to compute the two-body energy term.
For more details see Refs \cite{Bordbar80,big1}.

%%%%%%%%%%%%%%%%%%%%%%%%%%%%%%%%%%%%%%%%%%%%%%%%%%%%%%%%%%%%%%%%%%%%%%%%%%%%%%%%%%%%%%%%%%%%%%%%%%%%%%%%%%%%%%%%%%%%%%%%%%%%%%%%%%%%%%%%%%%%%

\section{Results and discussion }\label{NLmatchingFFtex}

In this paper, we study the liquid gas phase transition for the nuclear matter in the unpolarized, ferromagnetic and antiferromagnetic states.
It should be mentioned that the cases $\delta_p=\delta_n=0$,
$\delta_p=\delta_n$ and $\delta_p=-\delta_n$ are called the unpolarized, ferromagnetic and antiferromagnetic states, respectively \cite{big1,big2}. Ferromagnetic and antiferromagnetic states may also called polarized states.

\begin{figure}[th]
\center{\includegraphics{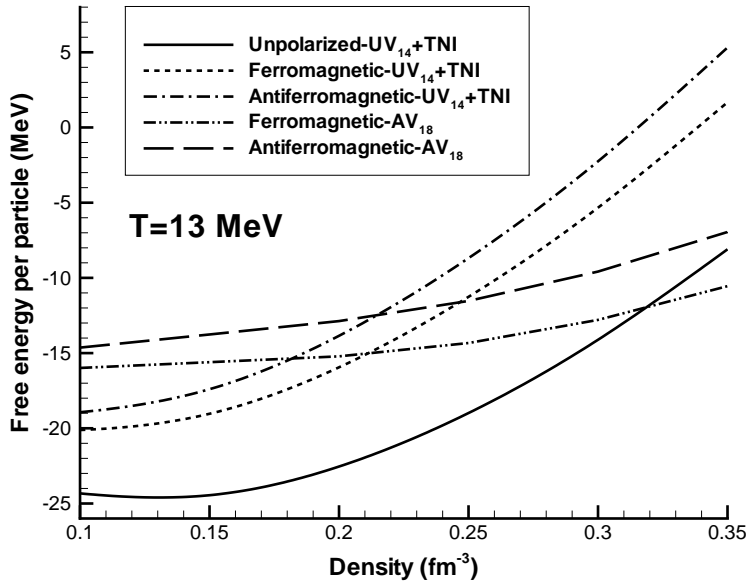}}
 \caption{The free energy per particle versus the density with $UV_{14}$+TNI and $AV_{18}$ potentials
for different magnetic states (with $\delta_p=\pm0.5$ for the ferromagnetic state and also  $\delta_p=\pm0.5$ for the antiferromagnetic state) at a fixed temperature.}
\label{E1}
\end{figure}

Fig. \ref{E1} shows the free energy per particle versus the density
with $UV_{14}$+TNI and $AV_{18}$ potentials for different magnetic states.
Obviously, at lower densities, the convexity of free energy which is the
condition of stability, is violated and nuclear matter is mechanically instable. This leads
to a first order liquid gas phase transition in nuclear matter. From Fig. \ref{E1}, it can be seen that at higher densities,
the convexity condition remains valid, and the system is stable. We can see that in the case of $UV_{14}$+TNI potential, the free energy
is more sensitive to the variation of density.

\begin{figure}[th]
\center{\includegraphics{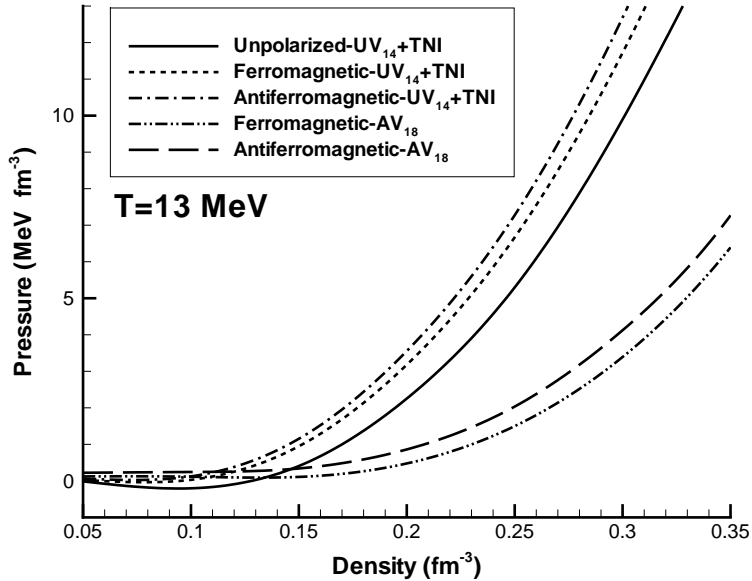}}
 \caption{Same as Fig. \ref{E1} but for the pressure.}
\label{1}
\end{figure}

Fig. \ref{1} presents the pressure-density isotherms at a fixed temperature
with $UV_{14}$+TNI and $AV_{18}$ potentials for three magnetic states. We see that for all magnetic states,
the isotherms corresponding to temperature $T=13\ MeV$ are below those of the corresponding critical temperatures.
At different magnetic states, the system shows a mechanical instability.
For both applied potentials, in the antiferromagnetic state, we see the stiffest
equation of state (EOS) for a fixed temperature.
This leads to the lower value of the critical temperature and critical density.
By comparing the isotherms, we realize that for unpolarized state,
the instable region is the most extensive. This indicates that for unpolarized system,
the second order phase transition occurs at a higher temperature.

\begin{figure}[th]
\center{\includegraphics{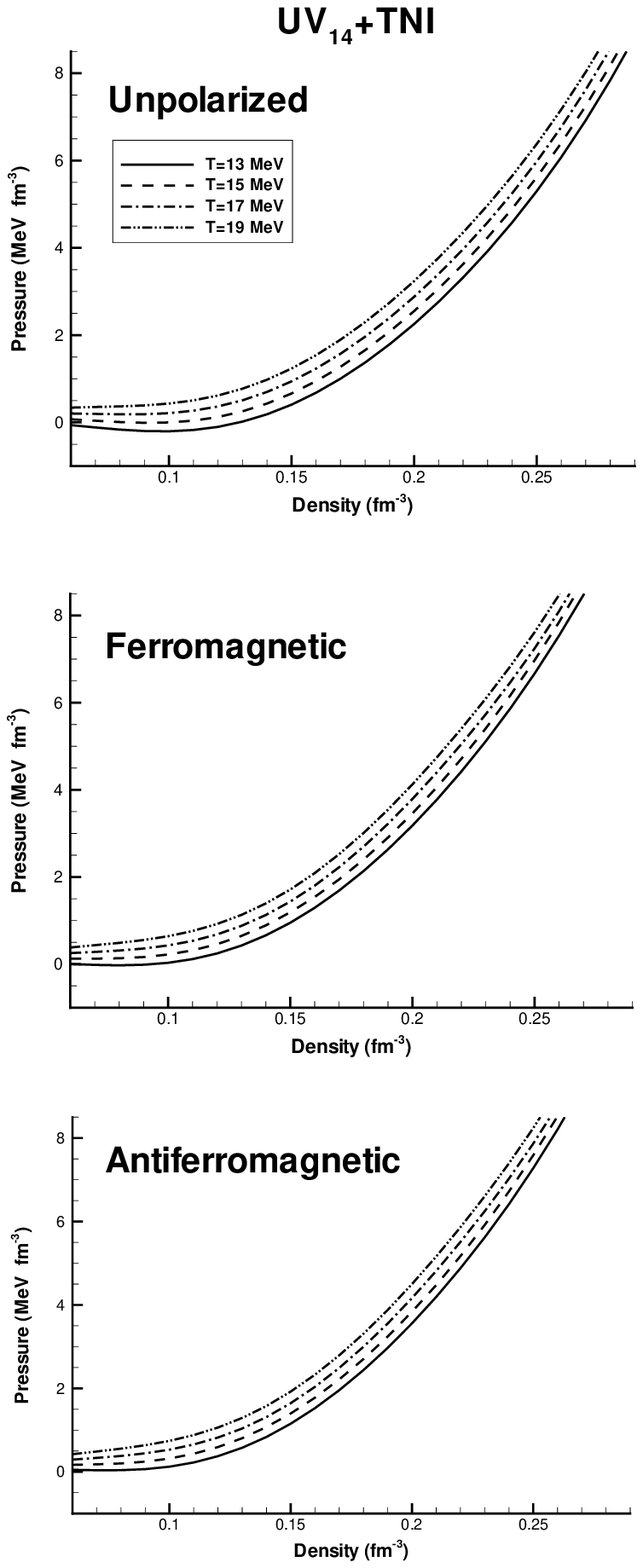}}
 \caption{The pressure-density isotherms for $UV_{14}$+TNI potential at different
temperatures for three magnetic states (with $\delta_p=\pm0.5$ for the polarized states).}
\label{2}
\end{figure}

\begin{figure}[th]
\center{\includegraphics{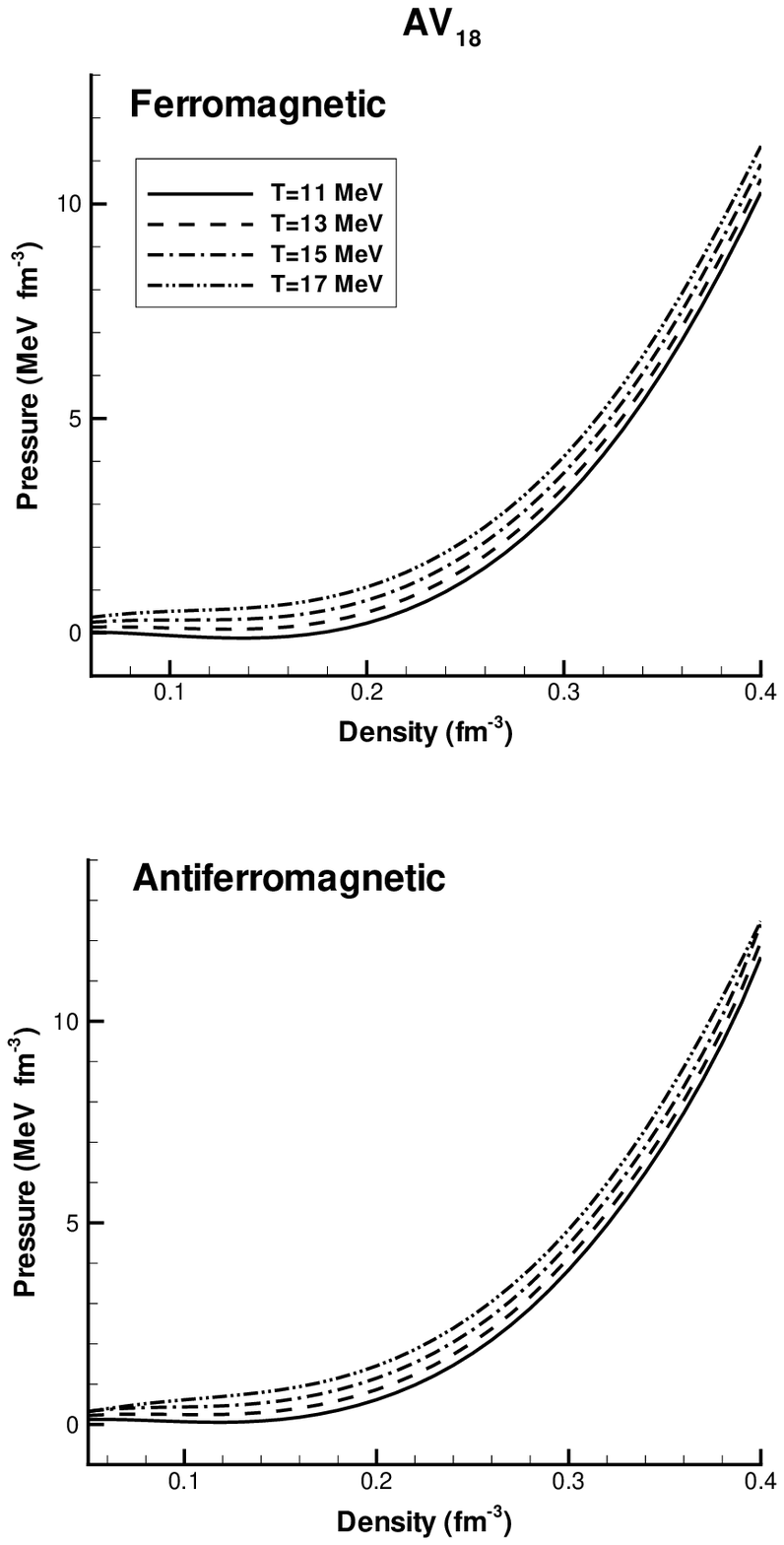}}
 \caption{Same as Fig. \ref{2} but for $AV_{18}$ potential.}
\label{3}
\end{figure}

Figs. \ref{2} and \ref{3} show the pressure-density isotherms at different
temperatures with $UV_{14}$+TNI and $AV_{18}$ potentials for three magnetic states.
The isotherms present a typical Van der Waals like behavior in which the liquid and
gaseous phases coexist. For all magnetic states, we have found that
the instable region reduces by increasing the temperature.
The properties of the mixed phase in thermal equilibrium can be calculated by
applying the equal-area Maxwell construction. To calculate the pressure of nuclear matter during the phase transition,
we employ the Maxwell construction  as shown in Figs.  \ref{6} and \ref{7}.
Between the left and right crossing points, the nuclear matter does not experience the unstable curve (the solid line), but physically it goes through
the curve at constant pressure (the dashed line). The stable states of nuclear matter between  the left and right crossing points correspond to the points on the dashed line. In fact, the continuous evolution with phase mixing appears instead of the discontinuity of the EOS known as the first order phase
transition.

\begin{figure}[th]
\center{\includegraphics[width=7.0cm]{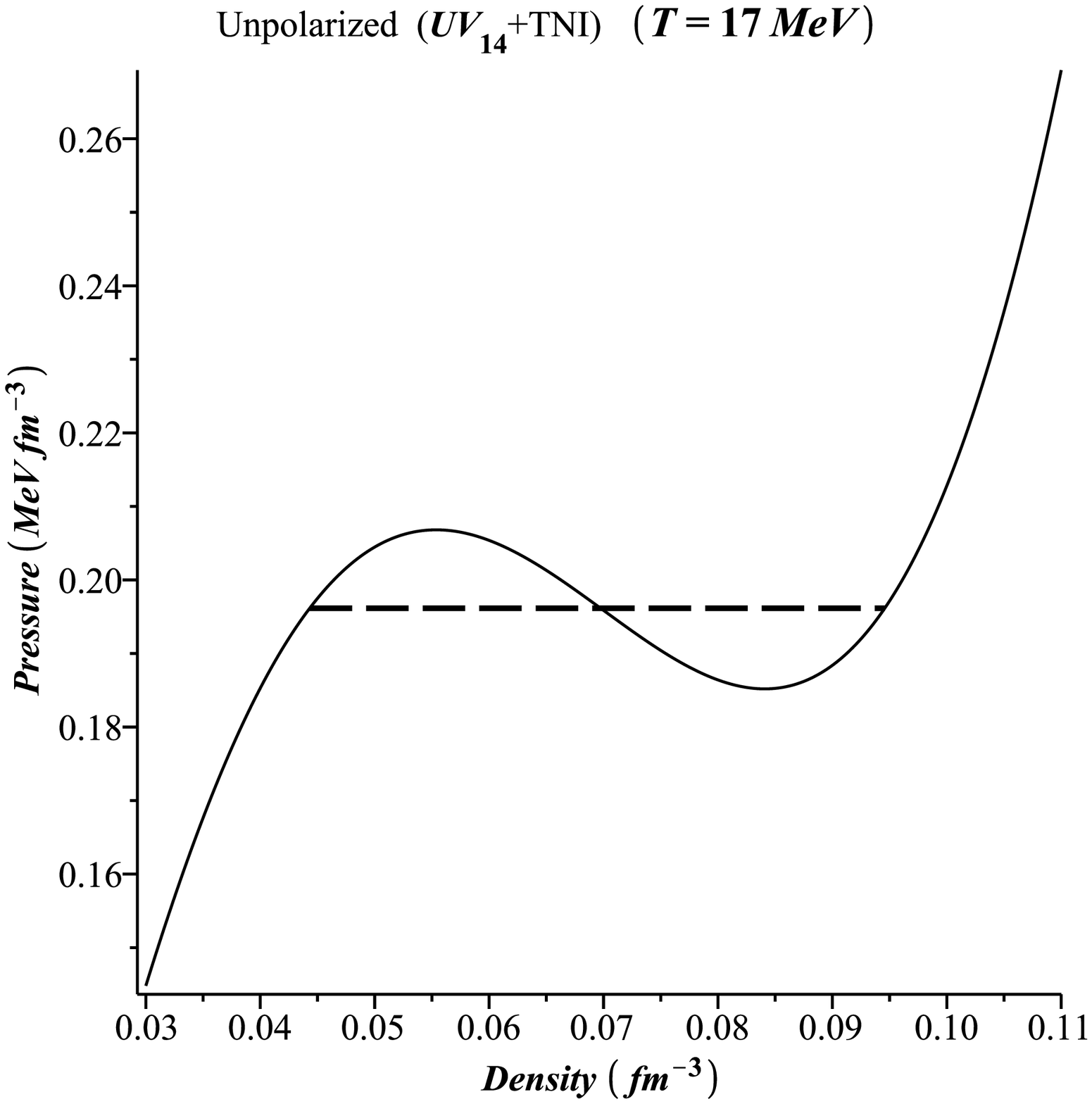}}
\center{\includegraphics[width=7.0cm]{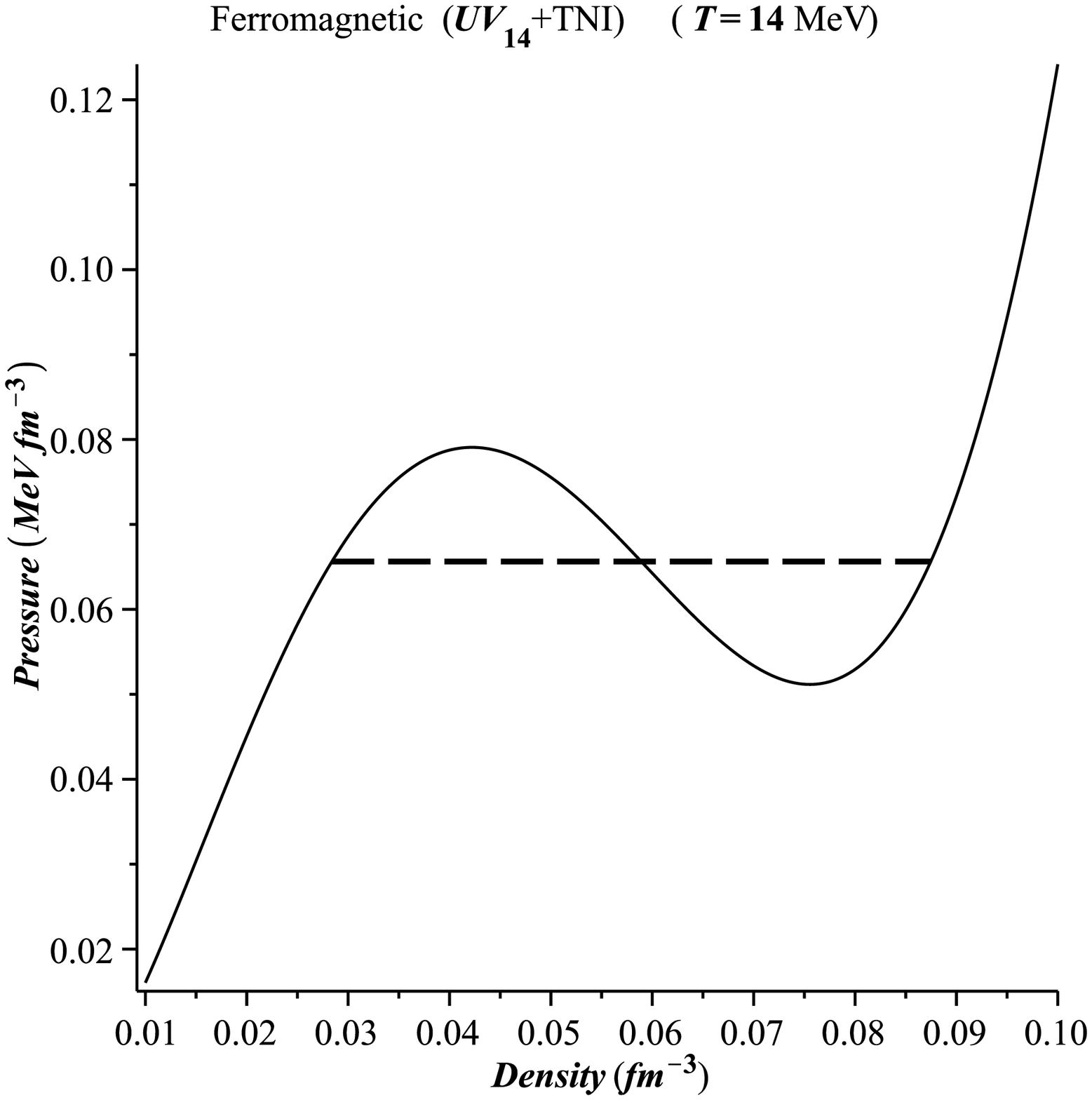}}
\center{\includegraphics[width=7.0cm]{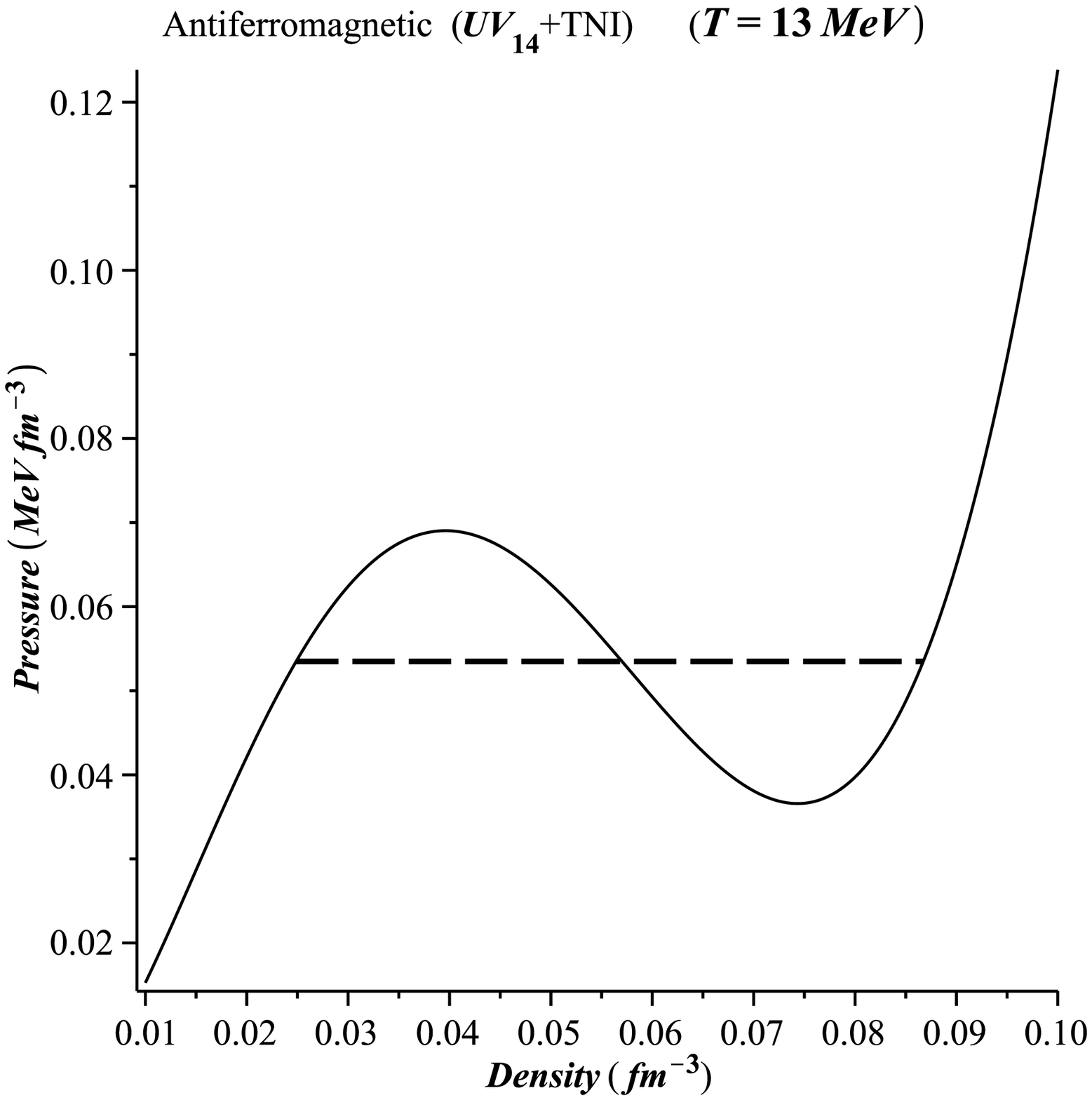}}
 \caption{The isotherms for $UV_{14}$+TNI potential at different magnetic states (with $\delta_p=\pm0.5$ for the polarized states). The
dashed line between  left and right crossing points was obtained by Maxwell construction.}
\label{6}
\end{figure}

\begin{figure}[th]
\center{\includegraphics[width=9.5cm]{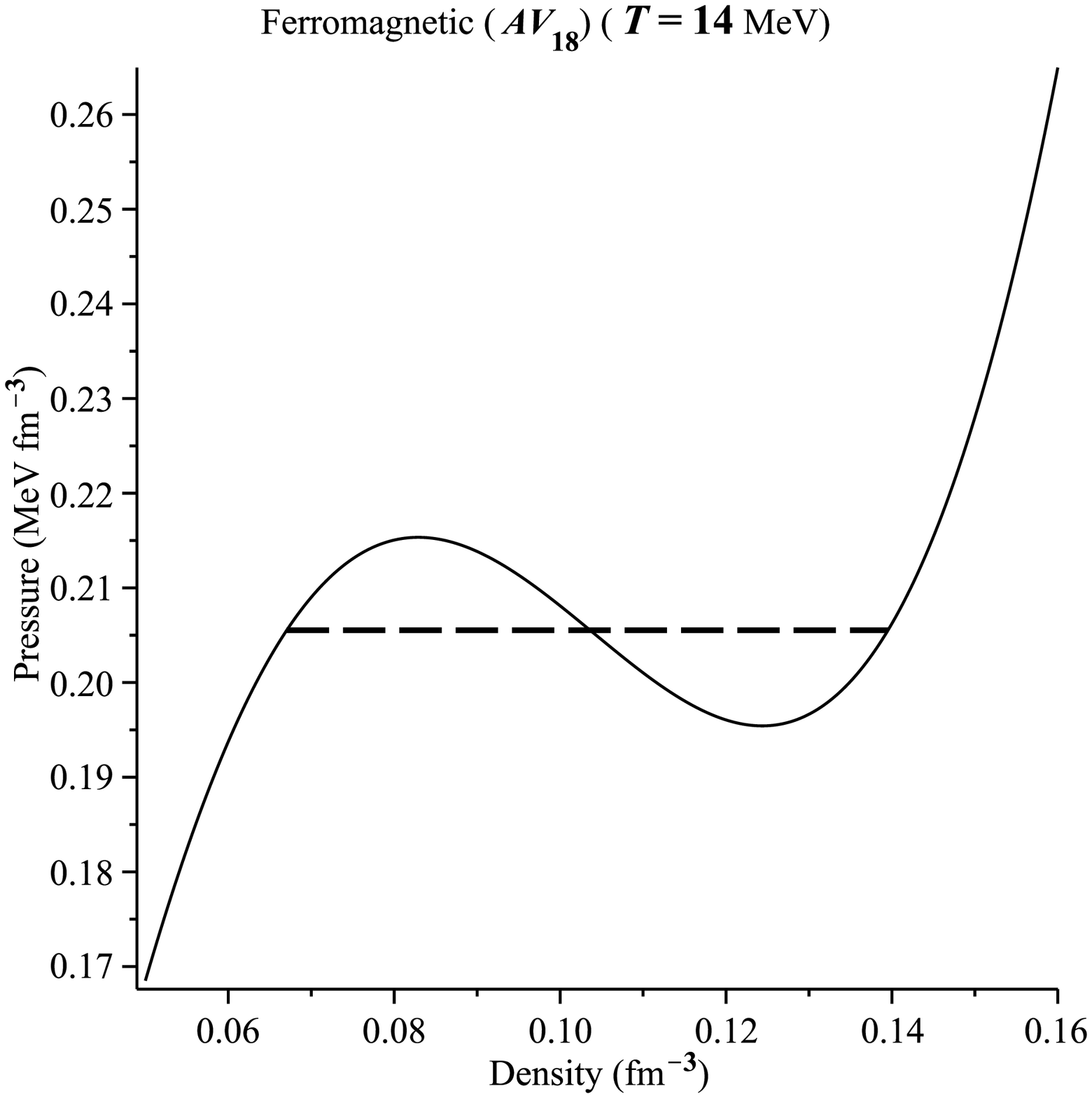}}
\center{\includegraphics[width=9.5cm]{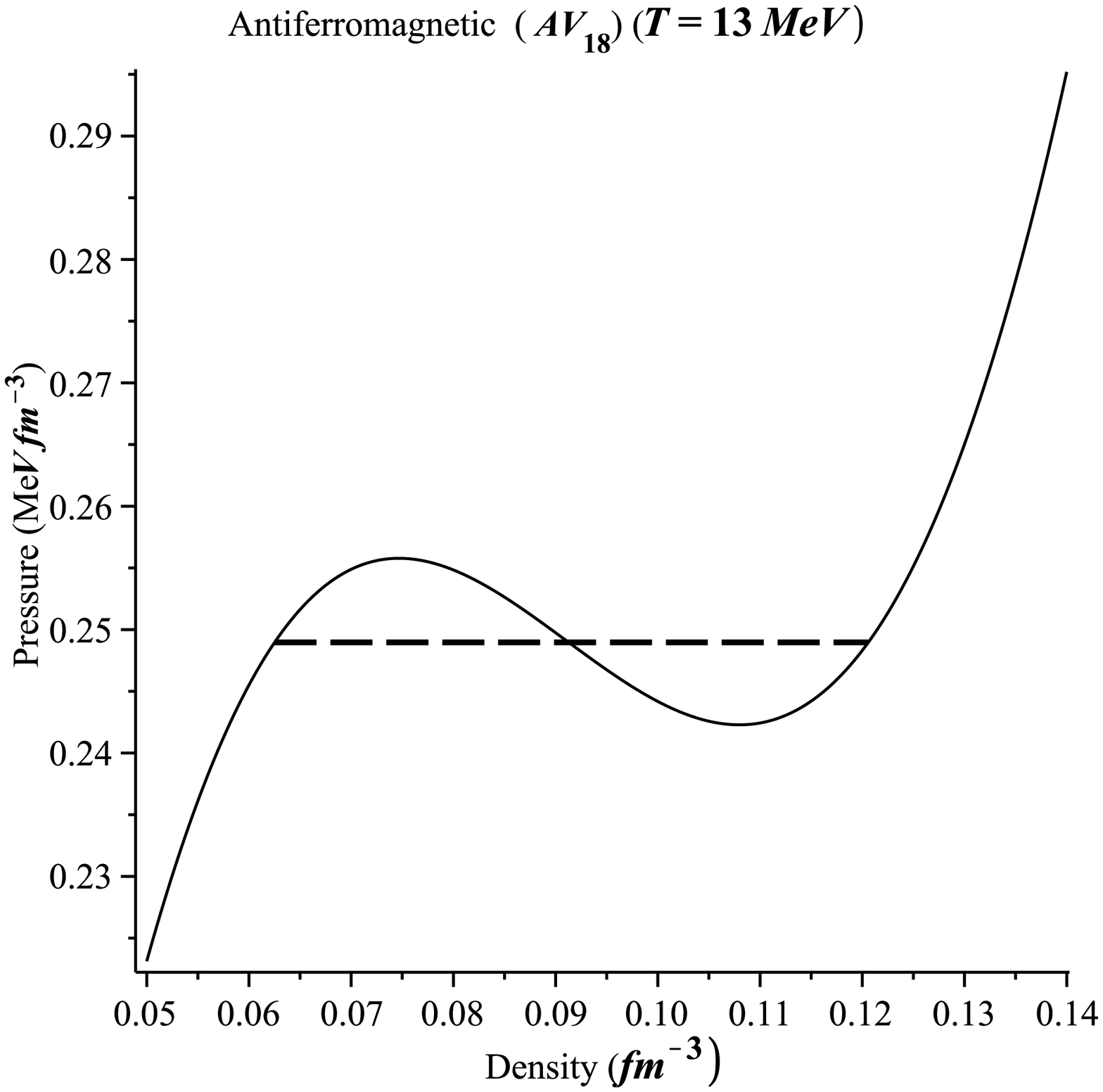}}
 \caption{Same as Fig. \ref{6} but for $AV_{18}$ potential.}
\label{7}
\end{figure}

\begin{figure}[th]
\center{\includegraphics{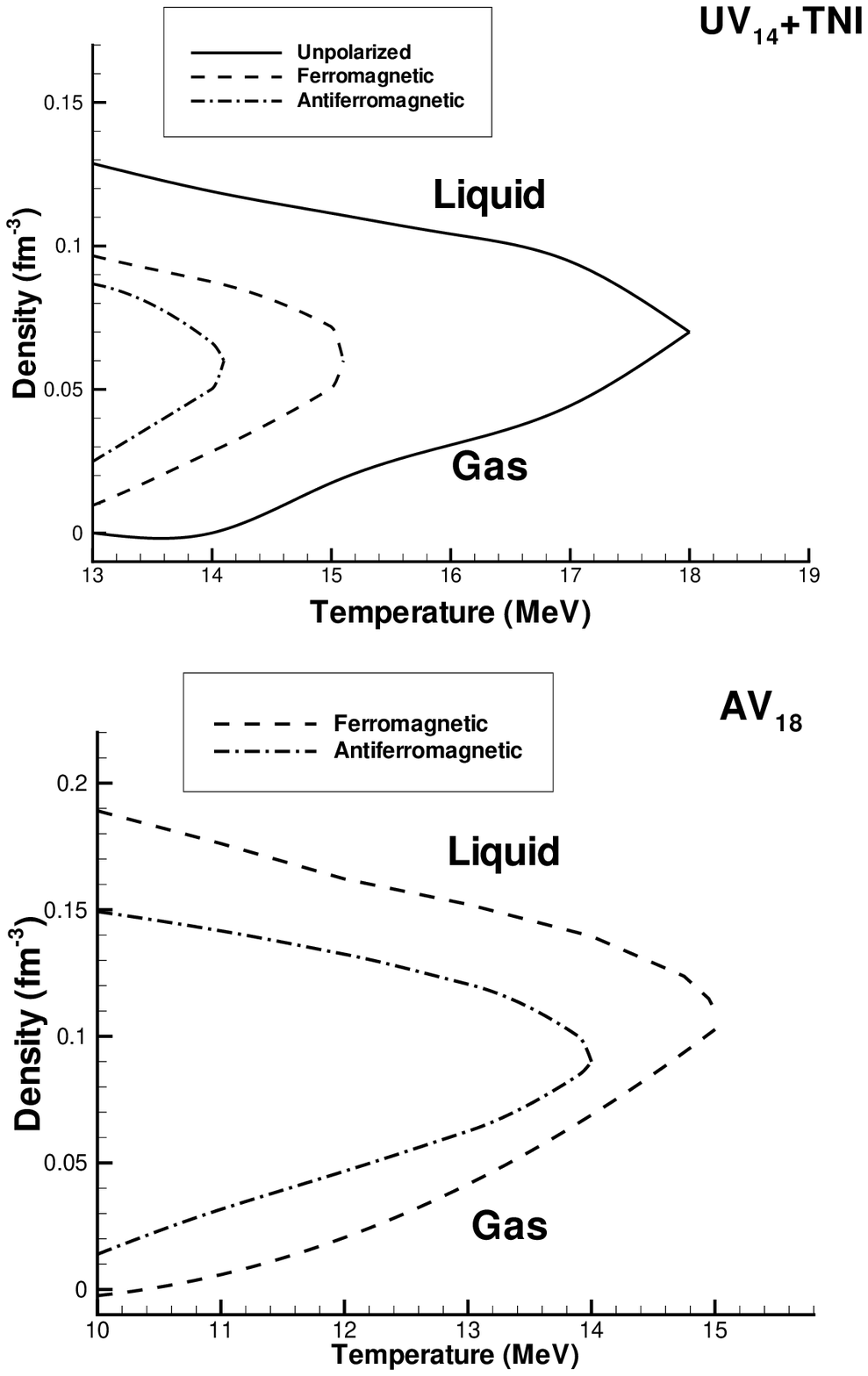}}
 \caption{The liquid gas coexistence curves for $UV_{14}$+TNI and $AV_{18}$ potentials at different magnetic states (with $\delta_p=\pm0.5$ for the polarized states).}
\label{8}
\end{figure}

Fig. \ref{8} shows the coexistence curves, the liquid and gas densities versus temperature, for $UV_{14}$+TNI and $AV_{18}$ potentials at different magnetic states. These curves are determined by the Maxwell construction.
As the temperature increases, the liquid density decreases and the gas density increases.
Inside the coexistence curve, the stable phase of nuclear matter is a mixture
of liquid and gas. The coexistence region decreases by increasing the temperature.
This region disappears at the critical temperature, and the densities become equal in which the nuclear matter experience a second order phase transition.
For polarized states at each temperature, the coexistence region is smaller than in the case of unpolarized state. In addition, for the antiferromagnetic state, the coexistence region is more limited than the ferromagnetic one. With two employed potentials, the coexistence region for the antiferromagnetic state disappears before the ferromagnetic one. This indicates that the critical temperature is lower for the antiferromagnetic state. At each temperature by comparing the coexistence regions obtained using two employed potentials,
 it is obvious that the coexistence region is wider for $AV_{18}$ potential.
At the critical temperature, $T_c$, with the
condition $(\frac{\partial P}{\partial \rho})
_{T,\delta_p}=(\frac{\partial^{2} P}{\partial \rho^{2}})
_{T,\delta_p}=0$,
the coexistence region changes to a point corresponding to the
critical pressure $P_c$ and critical density $\rho_c$. Our results for the critical temperature are presented in
Fig.  \ref{20}.
We can see that the critical temperature is a symmetric function of the spin polarization parameter.
This quantity decreases by increasing the magnitude of the spin polarization parameter.
The effect of spin polarization parameter on the critical temperature is more significant for  $AV_{18}$ potential.
Moreover, the effect of the spin polarization parameter on the critical temperature is more significant for the antiferromagnetic state.
At each value of the spin polarization parameter, the critical temperature for the ferromagnetic state is greater than the antiferromagnetic one.
The difference in the critical temperatures of the ferromagnetic and antiferromagnetic states is significant for larger values of polarization.
Using Fig.  \ref{20}, it is possible to extract the value of the spin polarization parameter at which the corresponding critical temperature is in agreement with the experimental value, $T_c\simeq16.6\ MeV$.
For the ferromagnetic nuclear matter with $UV_{14}$+TNI and $AV_{18}$ potentials, the critical temperatures are equal to $16.6\ MeV$ with $\delta_p=\pm0.31$ and $\delta_p=\pm0.44$, respectively. In addition, for the antiferromagnetic nuclear matter with $UV_{14}$+TNI and $AV_{18}$ potentials, the second order phase transition occurs at $16.6\ MeV$ when $\delta_p=\pm0.25$ and $\delta_p=\pm0.41$, respectively.

Our results for the critical density and pressure are presented in Tables~\ref{table2} and \ref{table3}.
Table~\ref{table2} indicates that the critical density for the case $\delta_p=\pm0.25$ is larger than that of $\delta_p=\pm 0.50$. In addition, with $AV_{18}$ potential, the critical density is more sensitive to the spin polarization parameter.
It should be noted that a similar behavior can be seen for the critical pressure from Table~\ref{table3}.

In the case of liquid gas phase transition, the
difference in the liquid and gas densities plays the role of order parameter. We define the order parameter as
 $m=\rho_{liquid}-\rho_{gas}$ to study the critical properties of the nuclear matter. This quantity is presented in Fig. \ref{9}
for the $UV_{14}$+TNI and $AV_{18}$ potentials at different magnetic states.
 It is obvious that the order parameter vanishes at the critical temperature.
 At each temperature, below the critical point, for the $UV_{14}$+TNI potential, the order parameter at the polarized states is smaller than that of the unpolarized state. It is evident that the order parameter for the ferromagnetic state is larger than the antiferromagnetic one.

Fig. \ref{10} presents the entropy per particle at the critical pressure as a function
of temperature. We can see that the entropy continues at the phase transition point, while as we will see, the heat capacity  diverges with a power law behavior at the phase transition point indicating that the
phase transition is of second order. At each temperature, the entropy of nuclear matter in antiferromagnetic state is larger than in
the ferromagnetic and unpolarized one.
The heat capacity at the critical pressure of the nuclear matter,
$ c_p=T(\frac{\partial S}{\partial T})_p$,
is plotted in Fig. \ref{11}.
This quantity  diverges with a power law behavior at the critical temperature confirming that the phase transition is the second order.

Fig. \ref{5} shows the isothermal
compressibility, $K_T=(\rho \frac{\partial P} {\partial \rho})^{-1}$,
as a function of temperature.
Below $T_c$, we have shown the liquid branch. For all temperatures,
the isothermal compressibility is also computed at $P_c$.
It is obvious that near the critical temperature from either side, the isothermal
compressibility diverges with a power law behavior. This phase transition is
specified by an order parameter which is non zero below the critical temperature
and is zero above it \cite{Eugene}, as we saw in Fig. \ref{9}. Close to the critical point, the fluctuations intervenes
dominate \cite{callen}.

\begin{figure}[th]
\center{\includegraphics[width=9.5cm]{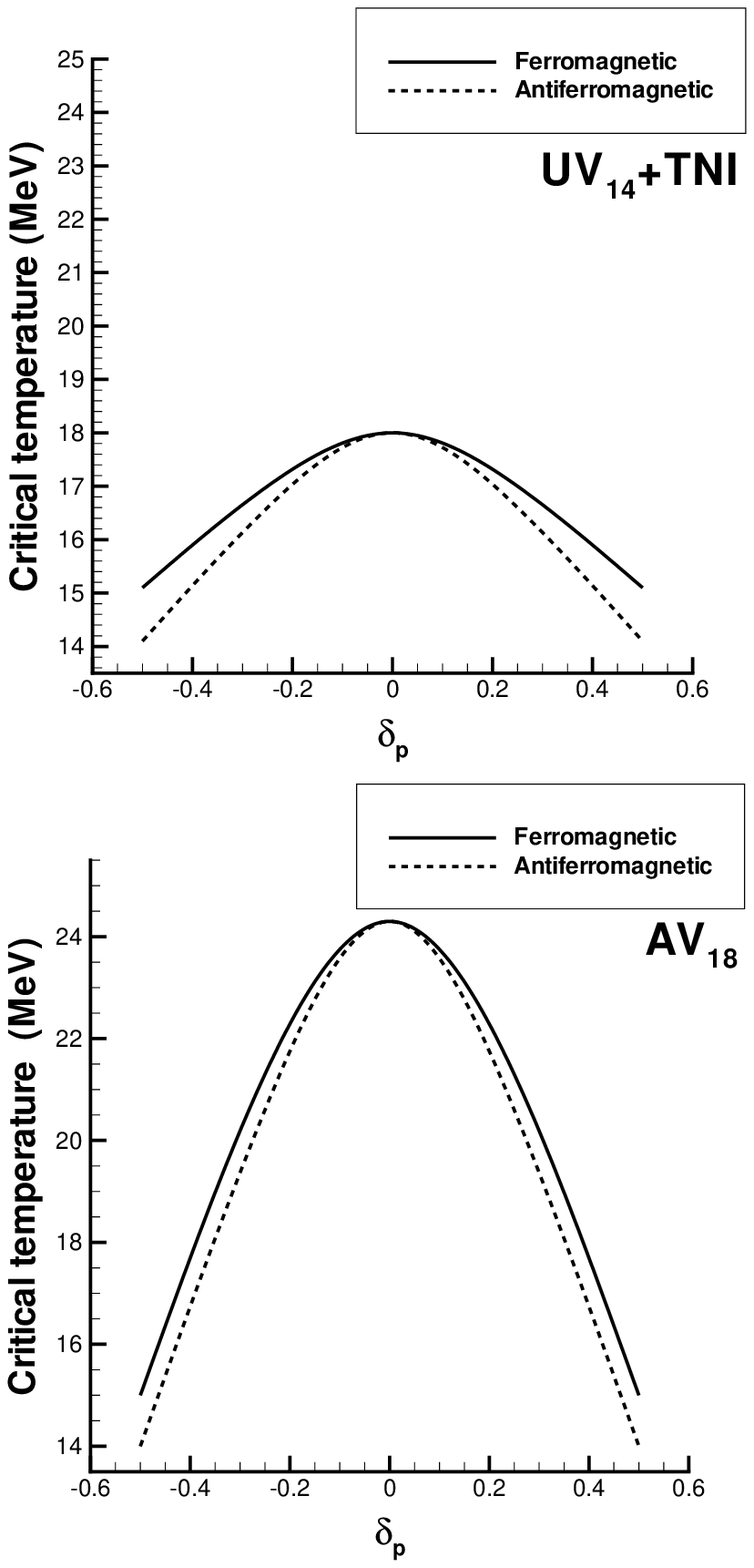}}
 \caption{The critical temperature versus the spin polarization parameter,  $\delta_p$,  for $UV_{14}$+TNI and $AV_{18}$ potentials at different magnetic states.}
\label{20}
\end{figure}

\begin{table}[pt]
\caption{Our results for the critical density of symmetric nuclear matter for $UV_{14}$+TNI and $AV_{18}$ potentials at different magnetic states.}
\label{table2}
\begin{center}  {\footnotesize
\begin{tabular}{|c|c|c|c|c|}
\hline Potential  & \multicolumn{1}{c|}{ Magnetic State} &
\multicolumn{1}{c|}{ $\delta_p$}&
\multicolumn{1}{c|}{$\rho_c\ (fm^{-3})$}\\\hline
$UV_{14}$+TNI   &Unpolarized & 0.00 & 0.07 \\
\cline{2-4}
 &Ferromagnetic &$\pm 0.25$ &0.07  \\\cline{3-4}
  & &$\pm 0.50$ & 0.06 \\
\cline{2-4}
  &Antiferromagnetic & $\pm0.25$ & 0.07 \\\cline{3-4}
    & & $\pm0.50$ & 0.06 \\\hline
$AV_{18}$\   &Ferromagnetic & $\pm0.25$ & 0.13  \\\cline{3-4}
  & &$ \pm0.50$ & 0.11 \\
\cline{2-4}
\  &Antiferromagnetic & $\pm0.25$  &0.13  \\\cline{3-4}
  & & $\pm0.50$ & 0.09 \\\hline
\end{tabular} }
\end{center}
\end{table}

\begin{table}[pt]
\caption{Same as Table~\ref{table2} but for the critical pressure.}
\label{table3}
\begin{center}  {\footnotesize
\begin{tabular}{|c|c|c|c|c|}
\hline Potential  & \multicolumn{1}{c|}{ Magnetic State} &
\multicolumn{1}{c|}{ $\delta_p$}&
\multicolumn{1}{c|}{$P_c\ (MeV  fm^{-3})$}\\\hline
$UV_{14}$+TNI   &Unpolarized & 0.00 &0.28  \\
\cline{2-4}
 &Ferromagnetic & $\pm0.25$ &0.21  \\\cline{3-4}
  & &$\pm 0.50$ &0.13  \\
\cline{2-4}
  &Antiferromagnetic & $\pm0.25$ & 0.19 \\\cline{3-4}
    & & $\pm0.50$ &0.11  \\\hline
$AV_{18}$\   &Ferromagnetic & $\pm0.25$ &  0.82 \\\cline{3-4}
  & & $\pm0.50$ & 0.30 \\
\cline{2-4}
\  &Antiferromagnetic & $\pm0.25$  & 0.76 \\\cline{3-4}
  & & $\pm0.50$ & 0.33 \\\hline
\end{tabular} }
\end{center}
\end{table}
%
%%%%%%%%%%%%%%%%%%%%%%%%%%%%%%%%%%%%%%%%%%%%%%%%%%%%%%%%%%%%%%%%%%%%%%%%%%%%%%%%%%%%

\begin{table}[pt]
\caption{Critical temperature for different potentials and methods (with $\delta_p=\pm0.5$ for our polarized states). }
\label{table5}
\begin{center}  {\footnotesize
\begin{tabular}{|c|c|c|c|c|}
\hline  \multicolumn{1}{|c|}{Potential} &
\multicolumn{1}{c|}{Method}&
\multicolumn{1}{c|}{$T_c\ (MeV)$}&
\multicolumn{1}{c|}{Reference}\\\hline
$UV_{14}$+TNI & LOCV & $18.0 $ &Our result(Unpolarized) \\
$UV_{14}$+TNI & LOCV & $15.1 $ &Our result(Ferromagnetic) \\
$UV_{14}$+TNI & LOCV & $14.1 $ &Our result(Antiferromagnetic) \\
$AV_{18}$ & LOCV & 15.0 &Our result(Ferromagnetic) \\
$AV_{18}$ & LOCV & 14.0 &Our result(Antiferromagnetic) \\
$AV_{18}$ & LOCV & 24.3 &\cite{Bordbar01} \\
$AV_{18}$ & SCGF  &11.6 &\cite{Rios}  \\
$AV_{18}$& BHF & 18.1  &\cite{Rios} \\
CDBONN & SCGF & 18.5 &\cite{Rios} \\
CDBONN & BHF & 23.3 &\cite{Rios} \\
$UV_{14}$+TNI & Variational Calculation & 17.5 &\cite{Friedman} \\
$AV_{14}$ & BlochDe Dominicis expansion & 21 &\cite{Baldo} \\
$AV_{14}$+TNI & BlochDe Dominicis expansion & 20 &\cite{Baldo} \\
$AV_{18}$+TNI & BHF & 13 &\cite{Zuo} \\
Experimental value& &$16.6$ &\cite{Natowitz}\\
\hline
\end{tabular} }
\end{center}
\end{table}

The critical temperature obtained using different potentials and methods have been compared
in Table~\ref{table5}. It can be seen that for $AV_{18}$ potential with SCGF method and
also for $AV_{18}+TNI$ potential with BHF method, the extracted critical temperatures are lower than those of our results. Among the results given by $UV_{14}+TNI$, $AV_{14}$ and $AV_{14}+TNI$ potentials with the
variational calculations and BlochDe Dominicis expansion, it is obvious that
the value $T_c\simeq21\ MeV$ corresponding to $AV_{14}$ and BlochDe Dominicis expansion
is almost close to our result, $T_c\simeq24.3\ MeV$ \cite{Bordbar01}. Our critical temperature for the unpolarized system, $T_c\simeq24.3\ MeV$,
is in a nearly good agreement with the obtained value  by the CDBONN potential in the BHF approximation ($T_c \simeq23.3\ MeV$).

%%%%%%%%%%%%%%%%%%%%%%%%%%%%%%%%%%%%%%%%%%%%%%%%%%%%%%%%%%%%%%%%%%%%%%%%%%%%%%%%%%%%%%%%%%%%%%%%%%%%%%%%%%%%%%%%%%%%%%%%%%%%

\section{Critical Exponents for Spin Polarized Nuclear Matter}
In this paper, we are interested in the critical exponents of spin polarized nuclear matter which describes the behavior of the thermodynamic properties of our system near the critical point.
We consider the critical isotherm,
\begin{eqnarray}
P-P_c\sim(\rho-\rho_c)^{\delta};    \rho \longrightarrow \rho_c,
 \end{eqnarray}
where $T=T_c$. The slope of this thermodynamic property on log-log scale corresponds to the critical exponent $\delta$. The values of this critical exponent with $UV_{14}$+TNI and $AV_{18}$ potentials for different magnetic states which are calculated using Fig. \ref{14} are presented in Table~\ref{table4}.

\begin{figure}[th]
\center{\includegraphics{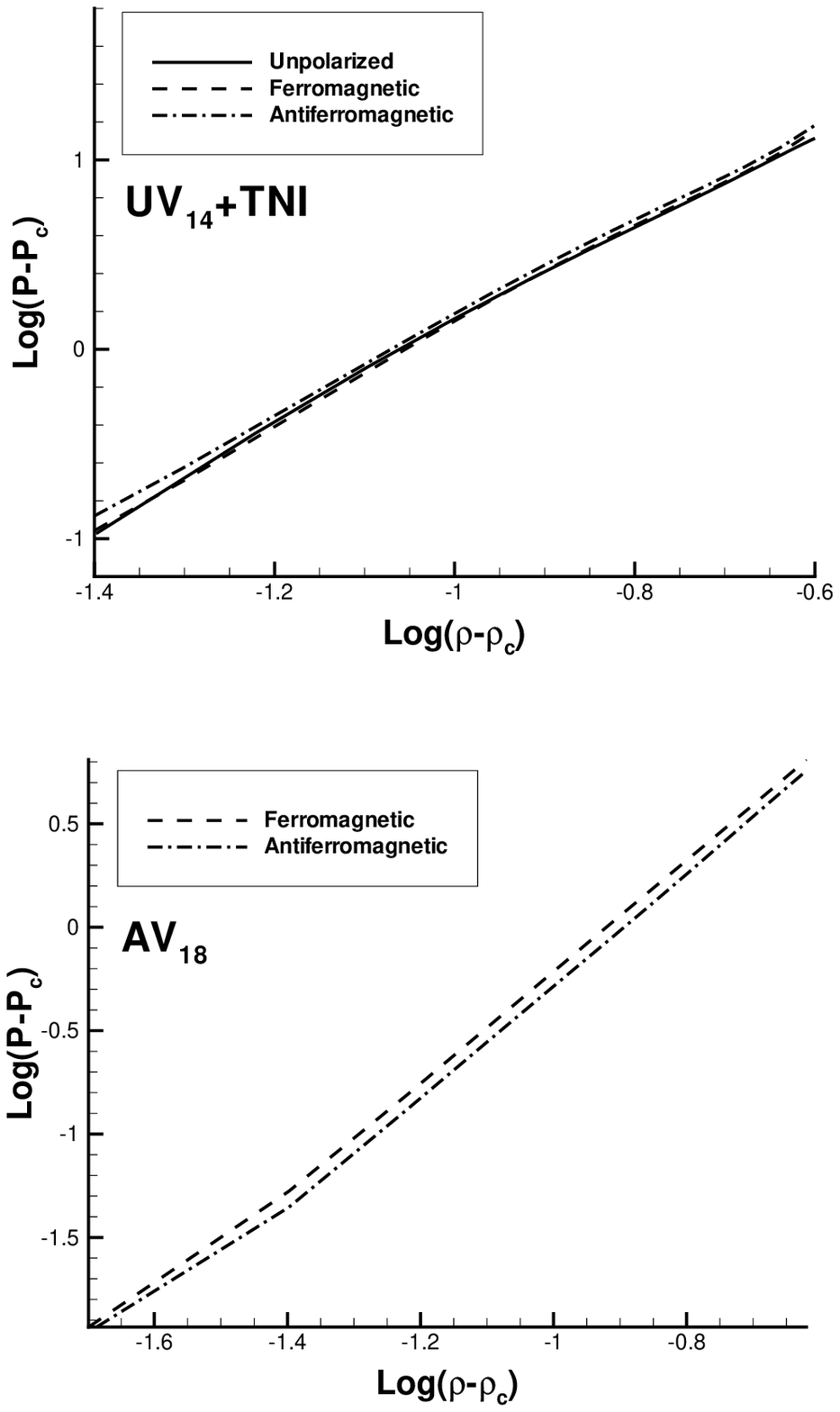}}
 \caption{The ($P-P_c$) versus ($\rho-\rho_c$) at critical temperature ($T_c$) on log-log scale with $UV_{14}$+TNI and $AV_{18}$ potentials for different magnetic states (with $\delta_p=\pm0.5$ for the polarized states).}
\label{14}
\end{figure}

\begin{figure}[th]
\center{\includegraphics[width=15.5cm]{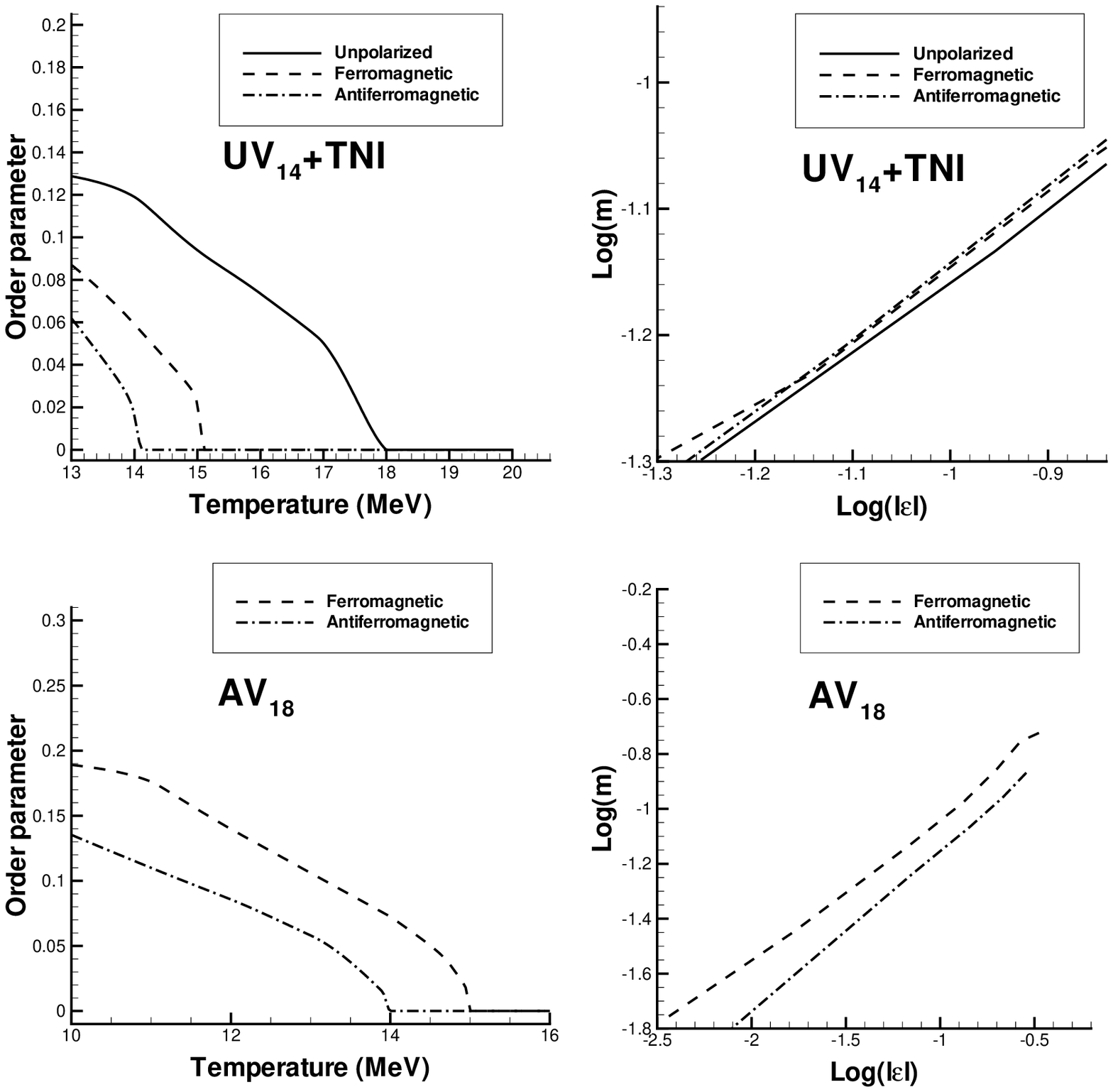}}
 \caption{Two left panels: The order parameter for the liquid gas phase transition as a function of temperature with $UV_{14}$+TNI and $AV_{18}$ potentials for different magnetic states (with $\delta_p=\pm0.5$ for the polarized states). Two right panels: The order parameter versus $\varepsilon$ on log-log scale.}
\label{9}
\end{figure}

 By  definition of
\begin{eqnarray}
 \varepsilon=\frac{T-T_c}{T_c},
 \end{eqnarray}
and considering the order parameter
as a function of $\varepsilon$ on the log-log scale in Fig. \ref{9}, we can calculate the exponent $\beta$ for this parameter,
\begin{eqnarray}
 m=(-\varepsilon)^{\beta}; \varepsilon \rightarrow 0.
 \end{eqnarray}
Our results for the value of $\beta$ with $UV_{14}$+TNI and $AV_{18}$ potentials for different magnetic states are presented in Table~\ref{table4}.

\begin{figure}[th]
\center{\includegraphics[width=10.5cm]{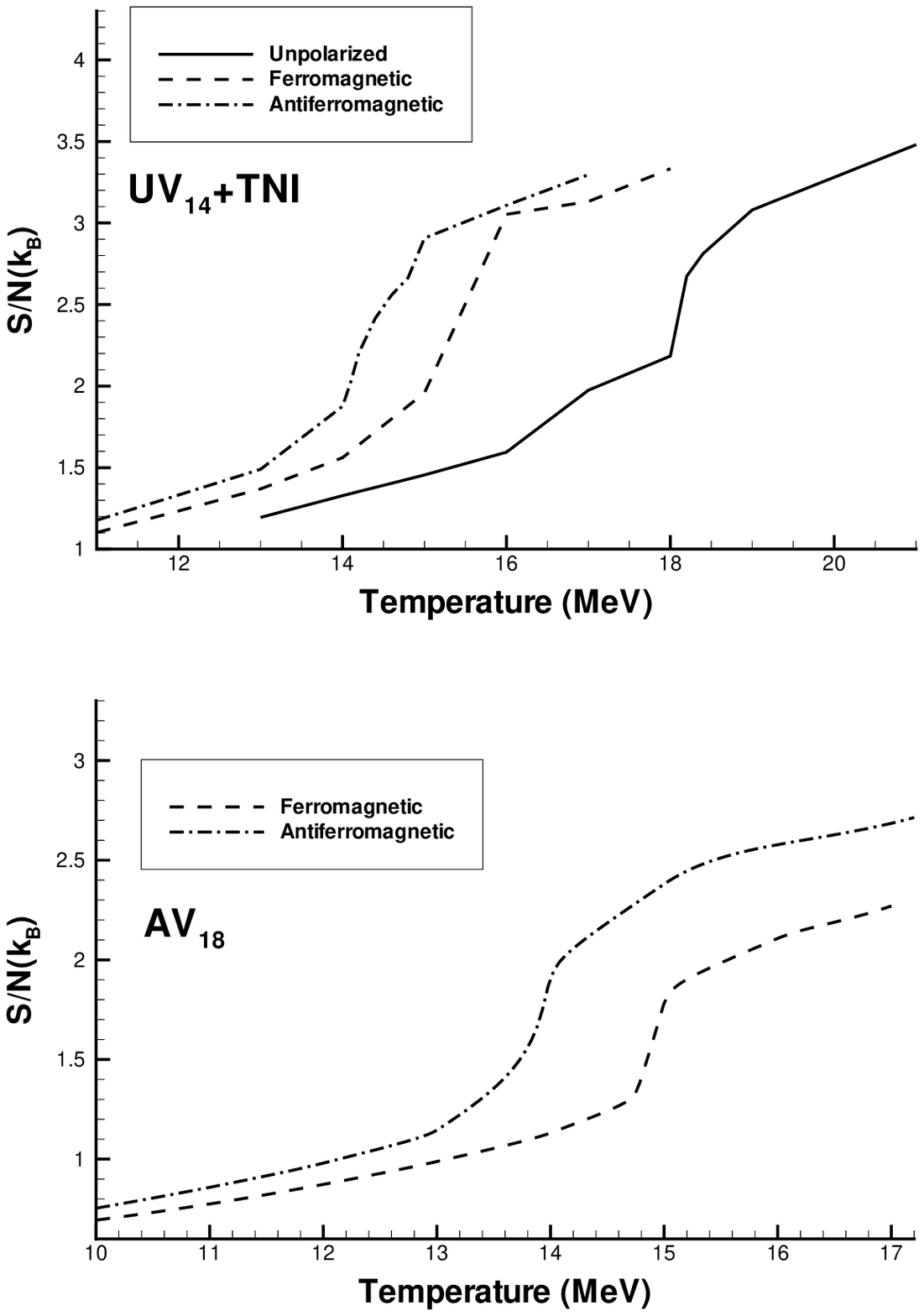}}
 \caption{The entropy per particle as a function of temperature at critical pressure with $UV_{14}$+TNI and $AV_{18}$ potentials for different magnetic states (with $\delta_p=\pm0.5$ for the polarized states).}
\label{10}
\end{figure}

The heat capacity is related to the exponent $\alpha$ by
\begin{eqnarray}
c_p=(-\varepsilon)^{-\alpha}; \varepsilon\rightarrow0.
 \end{eqnarray}
The values of the critical exponent $\alpha$ with $UV_{14}$+TNI and $AV_{18}$ potentials for different magnetic states which are calculated using Fig. \ref{11} are presented in Table~\ref{table4}.

\begin{figure}[th]
\center{\includegraphics[width=14.5cm]{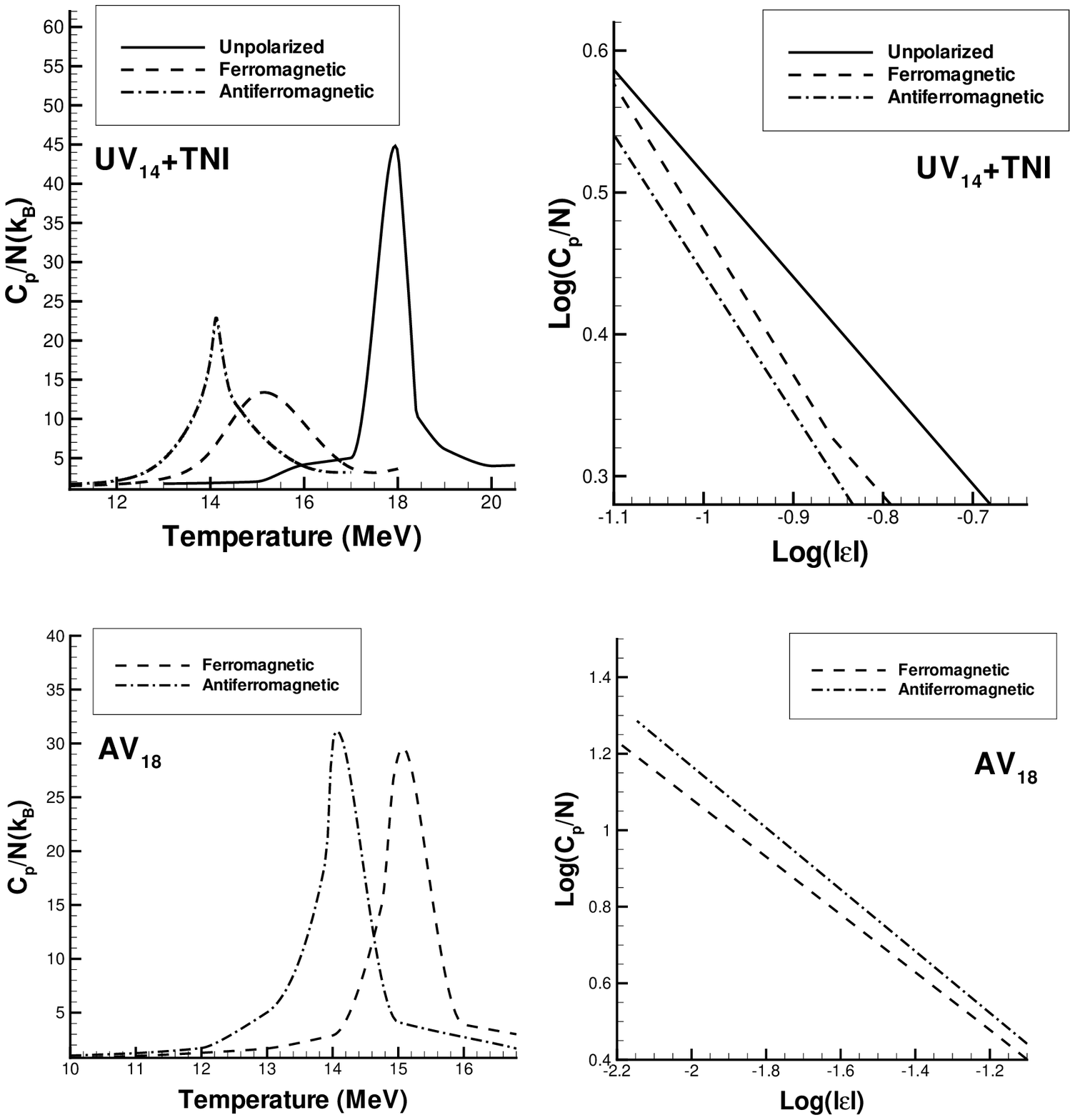}}
 \caption{Two left panels: The heat capacity per particle at critical pressure as a function of temperature with $UV_{14}$+TNI and $AV_{18}$ potentials for different magnetic states (with $\delta_p=\pm0.5$ for the polarized states). Two right panels: The heat capacity per particle at critical pressure versus $\varepsilon$ on log-log scale. }
\label{11}
\end{figure}

\begin{figure}[th]
\center{\includegraphics[width=14.5cm]{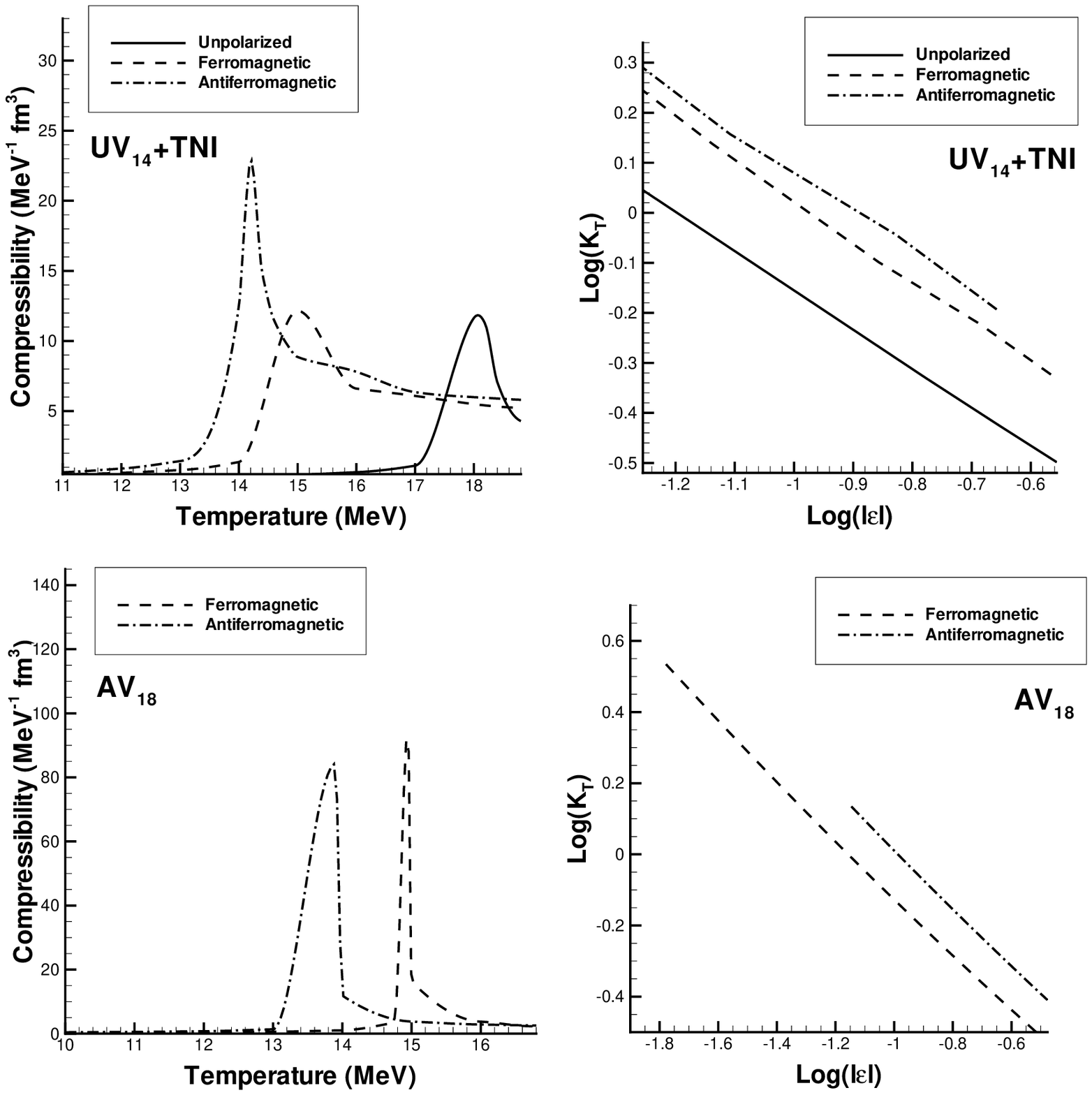}}
 \caption{Two left panels: The isothermal
compressibility as a function of temperature with $UV_{14}$+TNI and $AV_{18}$ potentials for different magnetic states (with $\delta_p=\pm0.5$ for the polarized states). Two right panels: The isothermal compressibility versus $\varepsilon$ on log-log scale.}
\label{5}
\end{figure}

To describe the isothermal compressibility near the critical
point, we calculate the critical exponent $\gamma$,
\begin{eqnarray}
K_T=(-\varepsilon)^{-\gamma}; \varepsilon\rightarrow 0.
 \end{eqnarray}
The values of this critical exponent which are obtained by Fig. \ref{5} are also presented in Table~\ref{table4}.
From Table~\ref{table4}, we can see that near the critical point, the critical exponents calculated with $UV_{14}$+TNI and $AV_{18}$ potentials for different magnetic states are nearly identical. However, there are some small differences in the values because of the numerical errors occurring in the calculations near the critical point.
From Table~\ref{table4}, it is evident that the Griffiths and Rushbrooke inequalities \cite{Huang,Griffiths}
\begin{eqnarray}
\alpha+2\beta+\gamma\geq2 ,
 \end{eqnarray}
 and
 \begin{eqnarray}
 \alpha+\beta(1+\delta)\geq2,
 \end{eqnarray}
are established for our critical exponents.
%%%%%%%%%%%%%%%%%%%%%%%%%%%%%%%%%%%%%%%%%%%%%%%%%%%%%%%%%%%%%%%%%%%%%%%%%%%%%
\begin{table}[pt]
\caption{The critical exponents for the symmetric nuclear matter with $UV_{14}$+TNI and $AV_{18}$ potentials for different magnetic states (with $\delta_p=\pm0.5$ for the polarized states).}
\label{table4}
\begin{center}  {\footnotesize
\begin{tabular}{|c|c|c|c|c|}
\hline Critical exponent  & \multicolumn{1}{c|} {Potential}  & \multicolumn{1}{c|}{Magnetic State} &
\multicolumn{1}{c|}{ }
\\\hline
$\beta$ &$UV_{14}$+TNI   &Unpolarized &$0.60078\pm0.02533$   \\
\cline{3-4}
& &Ferromagnetic   &$0.46184\pm0.01729$  \\\cline{3-4}
 & &Antiferromagnetic  &$0.55648\pm0.01592$  \\\cline{2-4}
&$AV_{18}$\   &Ferromagnetic & $0.53167\pm0.01598$  \\\cline{3-4}
\ & &Antiferromagnetic &$0.59037\pm0.00703$   \\\hline
$\delta$&$UV_{14}$+TNI   &Unpolarized & $2.54265\pm0.04097$  \\
\cline{3-4}
& &Ferromagnetic   & $2.65974\pm	0.03192$  \\\cline{3-4}
 & &Antiferromagnetic  & $2.59537	\pm0.02929$ \\\cline{2-4}
&$AV_{18}$\   &Ferromagnetic & $2.59047\pm	0.03184$  \\\cline{3-4}
\ & &Antiferromagnetic & $2.59117	\pm0.04159$  \\\hline
$\alpha$ &$UV_{14}$+TNI   &Unpolarized & $0.68480\pm	0.05970$  \\
\cline{3-4}
& &Ferromagnetic   & $0.66717\pm	0.03961$ \\\cline{3-4}
 & &Antiferromagnetic  &$0.60539	\pm0.04601$  \\\cline{2-4}
&$AV_{18}$\   &Ferromagnetic & $0.76092\pm0.00636$  \\\cline{3-4}
\ & &Antiferromagnetic & $0.81497\pm	0.01300$  \\\hline
$\gamma$&$UV_{14}$+TNI   &Unpolarized & $0.77925\pm	0.00334$   \\
\cline{3-4}
& &Ferromagnetic   & $0.87677\pm	0.01601$ \\\cline{3-4}
 & &Antiferromagnetic  & $0.86563\pm	0.02087$ \\\cline{2-4}
&$AV_{18}$\   &Ferromagnetic & $0.81733	\pm0.01216$  \\\cline{3-4}
\ & &Antiferromagnetic & $0.80605	\pm0.00806$  \\\hline
\end{tabular} }
\end{center}
\end{table}

\section{Summary and Conclusions}
In this work, we have calculated the thermodynamic properties of spin polarized
nuclear matter to study the liquid gas phase transition using
the lowest order constrained variational method employing the microscopic potentials.
The critical properties of unpolarized, ferromagnetic and antiferromagnetic nuclear matter
have been considered. By applying the Maxwell construction to study the mixed phase in thermal equilibrium, we have determined
the  pressure of nuclear matter during the phase transition and also the coexistence curve.
We have seen that the extension of the coexistence region is different for three magnetic states.
For unpolarized system, the second order
phase transition occurs at a higher temperature.
 In addition, for each value of the spin polarization parameter,
the critical temperature is greater for the ferromagnetic state.
Studying the critical temperature for different
spin polarization parameters, it has been shown that the critical temperature is a symmetric function of the spin polarization
parameter. Moreover, an increase in the magnitude of the spin
polarization parameter reduces the critical temperature.
 In investigation of the critical temperature, it has been found that
the antiferromagnetic nuclear matter is more sensitive to the value of the spin polarization parameter.
The critical temperature of our polarized nuclear matter with a specific value of spin polarization parameter is in a good agreement with the experimental result.
Studying the order parameter for the liquid gas phase transition,
 we have seen that below the critical point,
the order parameter for the unpolarized nuclear matter is larger than that of
polarized one.  It has been also clarified that the order parameter for the ferromagnetic state is larger
than the antiferromagnetic one.
 For our system, the heat capacity and isothermal compressibility
diverge with a power law behavior at the
phase transition point. This confirms that the phase transition is of second order.
Finally, we have calculated the critical exponents of the spin polarized nuclear
for different magnetic states, concluding that
the Griffiths and Rushbrooke inequalities
are established for these exponents.
\acknowledgements{
This work has been supported financially by the Center for Excellence in Astronomy and Astrophysics (CEAA-RIAAM).
G. H. Bordbar and Z. Rezaei wish to thank the Shiraz University Research Council.}

\end{document}